\documentclass[preprint,showpacs,preprintnumbers,amsmath,amssymb]{revtex4}
\usepackage{graphicx}
\usepackage{dcolumn}
\usepackage{bm}

\newcommand\br{\begin{eqnarray}}
\newcommand\er{\end{eqnarray}}
\newcommand\be{\begin{equation}}
\newcommand\ee{\end{equation}}

\newcommand\bc{\begin{center}}
\newcommand\ec{\end{center}}

\newcommand\PRL[3]{\textsl{Phys. Rev. Lett.} \textbf{#1}, #3 (#2)}

\newcommand\PRD[3]{\textsl{Phys. Rev.} \textbf{D#1}, #3 (#2)}

\newcommand\PLB[3]{\textsl{Phys. Lett.} \textbf{#1B}, #3 (#2)}
\newcommand\CQG[3]{\textsl{Class. Quantum Grav.} \textbf{#1}, #3 (#2)}

\newcommand\AoP[3]{\textsl{Ann. of Phys.} \textbf{#1}, #3 (#2)}
\newcommand\RMP[3]{\textsl{Rev. Mod. Phys.} \textbf{#1}, #3 (#2)}

\newcommand\IJMPA[3]{\textsl{Int. J. Mod. Phys.} \textbf{A#1}, #3 (#2)}
\newcommand\IJMPD[3]{\textsl{Int. J. Mod. Phys.} \textbf{D#1}, #3 (#2)}

\newcommand\JPA[3]{\textsl{J. Physics} \textbf{A#1}, #3 (#2)}

\newcommand\MPLA[3]{\textsl{Mod. Phys. Lett.} \textbf{A#1}, #3 (#2)}

\begin{document}


\title{Non Singular Origin of the Universe and its Present Vacuum Energy Density}

\author{E. I. Guendelman}
\email{guendel@bgu.ac.il} \affiliation{ Physics Department, Ben
Gurion University of the Negev, Beer Sheva 84105, Israel}

\bigskip

\begin{abstract}
We consider a non singular origin for the Universe starting from an Einstein static Universe, the so called "emergent universe" scenario, in the framework of a theory which uses two volume elements  $\sqrt{-{g}}d^{4}x$ and $\Phi d^{4}x$, where $\Phi $ is a metric independent density, used as an additional measure of integration. Also curvature, curvature square terms and for scale invariance a dilaton field $\phi$  are considered in the action. The first order formalism is applied.  The integration of the equations of motion associated with the
new measure gives rise to the spontaneous symmetry breaking (S.S.B) of scale invariance (S.I.).
After S.S.B. of S.I., it is found that a non trivial potential for the dilaton is generated. In the Einstein frame we also add a cosmological term that parametrizes the zero point fluctuations. The resulting effective potential for the dilaton contains two flat regions, for 
$\phi \rightarrow \infty$ relevant for the non singular origin of the Universe, followed by an inflationary phase and $\phi \rightarrow -\infty$, describing our present Universe. The dynamics of the scalar field becomes non linear and these non linearities
are instrumental in the stability of some of the emergent universe solutions, which exists for a parameter range
of values of the vacuum energy in $\phi \rightarrow -\infty$, which must be positive but not very big, avoiding the extreme fine tuning required to keep the vacuum energy density of the present universe small. Zero vacuum energy density for the present universe defines the threshold for the creation of the universe. 
\end{abstract}


\maketitle
\section{Introduction}
One  of the most important and intriging issues of modern physics is the so called "Cosmological Constant Problem" \cite{CCP1}, \cite{CCP2},\cite{CCP3}
(CCP), most easily seen by studying the apparently uncontrolled behavior of the zero point energies, which would lead to a corresponding
equally uncontrolled vacuum energy or cosmological constant term. Even staying at the classical level, the observed very small cosmological term in the present universe is still very puzzling.

One point of view to the CCP that has been popular has been to provide a bound based on the "anthropic principle" \cite{Anthropic}. In this approach, a too large Cosmological Constant will not provide the necessary conditions required for the existence of life, the anthropic principle
provides then an upper bound on the cosmological constant. 

One problem with this approach is for example that it relies on our knowledge of life as we know it and ignores the possibility that other life forms could be possible, for which other (unknown) bounds would be relevant, therefore the reasoning appears by its very nature subjective, since of course if the observed cosmological constant will be different, our universe will be different and this could include different kind of life that may be could have adjusted itself to a higher cosmological constant of the universe. But even accepting the validity of anthropic considerations, we still do not understand why the observed vacuum energy density must be positive instead of possibly a very small negative quantity. Accepting the anthropic explanation means may be also giving up on discovering important physics related to the  CCP and this may be the biggest objection.

Nevertheless, the idea of associating somehow restrictions on the origin of the universe with the cosmological constant problem seems interesting.
We will take on this point of view, but leave out the not understood concept of life out from our considerations. Instead, we will require, in a very specific framework, the non singular origin of the universe. The advantage of this point of view is that it is formulated in terms of ideas of physics alone, without reference to biology, which unlike physics, has not reached the level of an exact science. Another interesting consequence is that we can learn that a non singularly created universe may not have a too big cosmological constant, an effect that points to 
a certain type of gravitational suppresion of UV divergences in quantum field theory.

In this respect, one should point out that even in the context of the inflationary scenario \cite{Inflation1}, \cite{Inflation2}, \cite{Inflation3}, \cite{Inflation4}, which solves many cosmological problems, one still encounters the initial
singularity problem which remains unsolved, showing that the universe necessarily had a singular beginning
for generic inflationary cosmologies \cite{singularities1}, \cite{singularities2}, \cite{singularities3}, \cite{singularities4}, \cite{singularities5}.

Here  we will adopt the very attractive "Emergent Universe" scenario, where those conclusions concerning singularities can be avoided \cite{emerging1},\cite{emerging2}, \cite{emerging3}, \cite{emerging4},
\cite{emerging5}, \cite{emerging6}, \cite{emerging7}, \cite{emerging8}.
The way to escape the singularity in these models is to violate the geometrical assumptions of these theorems,
which assume i) that the universe has open space sections ii) the Hubble expansion is always greater than zero in the
past. In \cite{emerging1},\cite{emerging2} the open space section condition is violated since closed Robertson Walker
universes with $k=1$ are considered and the Hubble expansion can become zero, so that both i) and ii) are avoided.

In \cite{emerging1}, \cite{emerging2} even models based on standard General Relativity, ordinary matter and minimally coupled scalar fields
were considered and can provide indeed a non singular (geodesically complete) inflationary universe, with a past eternal Einstein
static Universe that eventually evolves into an inflationary Universe.

Those most simple models suffer however from instabilities, associated with the instability of the Einstein static universe.
The instability is possible to cure by going away from  GR, considering non perturbative corrections to the Einstein`s field
equations in the context of the loop quantum gravity\cite{emerging3}, a brane world cosmology \cite{emerging4}, considering the
Starobinski model for radiative corrections (which cannot be derived from an effective action)\cite{emerging5} or exotic matter\cite{emerging6}. In addition to this, the consideration of a Jordan Brans Dicke model also can provide
a stable initial state for the emerging universe scenario \cite{emerging7}, \cite{emerging8}.

In this paper we propose a different theoretical framework
where such emerging universe scenario is realized in a natural way, where instabilities are avoided and a succesfull inflationary phase
with a gracefull exit can be achieved. The  model we will use was studied first in \cite{SIchile}, however, we differ with \cite{SIchile} in our choice of the state (with a lower vacuum energy density) that best represents the present state of the universe. This is crucial, since as it should be obvious, the discussion of the CCP depends crucially on what vacuum we take. We will express the stability and existence conditions  for the non singular universe in terms of the energy of the vacuum of  our candidate for the present Universe. We will also by the  way discuss and correct a few typos in \cite{SIchile} and improve a bit the discussion of some notions discussed there as well. 

We work in the context of a theory built along the lines of the two measures theory (TMT) \cite{TMT1a}-\cite{TMT1r}, \cite{TMT2}, 
\cite{TMT3a}-\cite{TMT3c}, \cite{TMT4a}-\cite{TMT4e}, \cite{TMT5}
and more specifically in the context of the scale invariant
realization of such theories  \cite{TMT2}, \cite{TMT3a}-\cite{TMT3e}, \cite{TMT4a}-\cite{TMT4e}, \cite{TMT5}. These theories
can provide a new approach to  the cosmological constant problem and can be generalized to obtain also
a theory with a dynamical spacetime \cite{dyn} . We will consider a slight generalization of the TMT case, where,
we consider also the possible effects of zero point energy densities, thus "softly breaking" the basic structure of TMT for this purpose. We will show how the stated goals of a stable emerging universe
can be achieved in the framework of the model and also how the stability of the emerging universe imposes interesting constraints
on the energy density of the ground state of the theory as defined in this paper: it must be positive but not very large, thus the vacuum energy and therefore the term that softly breaks the TMT structure appears to be naturally controlled.

The paper will be organized as follows: First we review the principles
of the TMT and in particular the model studied in \cite{TMT2}, which has
global scale invariance and how this can be the basis for the emerging universe. Such model gives rise, in the effective
Einstein frame, to an effective potential for a dilaton field (needed to
implement an interesting model with global scale invariance) which has
a flat region. Following this, we
look at the generalization of this model \cite{TMT5} by adding a curvature square or simply "$R^{2}$ term"
and show that the resulting model contains now two flat
regions. The existence of two flat regions for the potential
is shown to be consequence of the s.s.b. of the scale symmetry.  We then consider the incorporation in the model of the zero point fluctuations, parametrized by a cosmological constant in the Einstein frame.  In this resulting 
model, there are two possible types of emerging universe solutions,
for one of those, the initial Einstein Universe can be stabilized due to the nonlinealities of the model, provided the vacuum energy 
density of the ground state is positive but not very large. This is a very satisfactory results, since it means that the stability
of the emerging universe prevents the vacuum energy in the present universe from being very large!. The transition from the emergent universe to the ground state goes through an intermediate inflationary phase, therefore reproducing the basic standard cosmological model as well.
We end with a discussion section and present the point of view that the creation of the universe can be considered as a "threshold event" for zero present vacuum energy density, which naturally gives a positive but small vacuum energy density.

\section{Introducing a new measure}

The general structure of general coordinate invariant theories is taken usually as
\begin{equation}\label{1}
S_{1} = \int{L_{1}}\sqrt{-g} d^{4}x 
\end{equation}
where $g =  det (g_{\mu\nu})$.  The introduction of $\sqrt{-g}$ is required since $d^{4}x$ by itself is not a scalar but the product
$\sqrt{-g} d^{4} x$ is a scalar. Inserting $\sqrt{-g}$,
which has the transformation properties of a
density, produces a scalar action $S_{1}$, as defined by eq.(\ref{1}), provided $L_{1}$ is a scalar.

    In principle nothing prevents us from considering other densities instead of
$\sqrt{-g}$. One costruction of such alternative "measure of integration", is obtained as follows:
 given 4-scalars $\varphi_{a}$ (a =
1,2,3,4), one can construct the density
\begin{equation}\label{2}
\Phi =  \varepsilon^{\mu\nu\alpha\beta}  \varepsilon_{abcd}
\partial_{\mu} \varphi_{a} \partial_{\nu} \varphi_{b} \partial_{\alpha}
\varphi_{c} \partial_{\beta} \varphi_{d}
\end{equation}
and consider in addition to the action $S_{1}$, as defined by eq.(\ref{1}),$S_{2}$, defined as
\begin{equation}\label{3}
S_{2} =  \int L_{2} \Phi d^{4} x
\end{equation}
$L_{2}$ is again some scalar, which may contain the curvature (i.e. the
gravitational contribution) and a matter contribution, as it can be the case for $S_{1}$, as defined by eq.(\ref{1}).

    In the action $S_{2}$ defined by eq.(\ref{3}) the measure carries degrees of freedom
independent of that of the metric and that of the matter fields. The most
natural and successful formulation of the theory is achieved when the
connection is also treated as an independent degree of freedom. This is
what is usually referred to as the first order formalism.

    One can consider both contributions, and allowing therefore both geometrical
objects to enter the theory and take as our action
\begin{equation}\label{e6}
S = \int L_{1} \sqrt{-g}d^{4}x + \int L_{2} \Phi  d^{4} x    
\end{equation}

 Here $L_{1}$ and
$L_{2}$ are
$\varphi_{a}$  independent.

    We will study now the dynamics of a scalar field $\phi$ interacting
with gravity as given by the following action, where except for the potential terms
$U$ and $V$ we have conformal invariance, the potential terms
$U$ and $V$ break down this to global scale invariance.

\begin{equation}\label{e9}
S_{L} =    \int L_{1} \sqrt{-g}   d^{4} x +  \int L_{2} \Phi d^{4} x 
\end{equation}
\begin{equation}\label{e10}
L_{1} = U(\phi)
\end{equation}

\begin{equation}\label{e11}
L_{2} = \frac{-1}{\kappa} R(\Gamma, g) + \frac{1}{2} g^{\mu\nu}
\partial_{\mu} \phi \partial_{\nu} \phi - V(\phi)
\end{equation}
\begin{equation}\label{e12}
R(\Gamma,g) =  g^{\mu\nu}  R_{\mu\nu} (\Gamma) , R_{\mu\nu}
(\Gamma) = R^{\lambda}_{\mu\nu\lambda}
\end{equation}
\begin{equation}\label{e13}
R^{\lambda}_{\mu\nu\sigma} (\Gamma) = \Gamma^{\lambda}_
{\mu\nu,\sigma} - \Gamma^{\lambda}_{\mu\sigma,\nu} +
\Gamma^{\lambda}_{\alpha\sigma}  \Gamma^{\alpha}_{\mu\nu} -
\Gamma^{\lambda}_{\alpha\nu} \Gamma^{\alpha}_{\mu\sigma}.
\end{equation}

The suffix $L$ in $S_{L}$ is to emphasize that here the curvature appears
only linearly. Here,  except for the potential terms
$U$ and $V$ we have conformal invariance, the potential terms
$U$ and $V$ break down this to global scale invariance. Since the breaking of local 
conformal invariance is only through potential terms, we call this a "soft breaking".

    In the variational principle $\Gamma^{\lambda}_{\mu\nu},
g_{\mu\nu}$, the measure fields scalars
$\varphi_{a}$ and the "matter" - scalar field $\phi$ are all to be treated
as independent
variables although the variational principle may result in equations that
allow us to solve some of these variables in terms of others.

For the case the potential terms
$U=V=0$ we have local conformal invariance

\begin{equation}\label{e14}
g_{\mu\nu}  \rightarrow   \Omega(x)  g_{\mu\nu}
\end{equation}

and $\varphi_{a}$ is transformed according to
\begin{equation}\label{e15}
\varphi_{a}   \rightarrow   \varphi^{\prime}_{a} = \varphi^{\prime}_{a}(\varphi_{b})
\end{equation}

\begin{equation}\label{e16}
\Phi \rightarrow \Phi^{\prime} = J(x) \Phi    
\end{equation}
 where $J(x)$  is the Jacobian of the transformation of the $\varphi_{a}$ fields.

This will be a symmetry in the case $U=V=0$ if 
\begin{equation}\label{e17}
\Omega = J
\end{equation}
Notice that $J$ can be a local function of space time, this can be arranged by performing for the 
$\varphi_{a}$ fields one of the (infinite) possible diffeomorphims in the internal $\varphi_{a}$ space.

 We can still retain a  global
scale invariance in model for very special exponetial form for the $U$ and $V$ potentials. Indeed, if we perform the global
scale transformation ($\theta$ =
constant)
\begin{equation}\label{e18}
g_{\mu\nu}  \rightarrow   e^{\theta}  g_{\mu\nu}
\end{equation}
then (9) is invariant provided  $V(\phi)$ and $U(\phi)$ are of the
form  \cite{TMT2}
\begin{equation}\label{e19}
V(\phi) = f_{1}  e^{\alpha\phi},  U(\phi) =  f_{2}
e^{2\alpha\phi}
\end{equation}
and $\varphi_{a}$ is transformed according to
\begin{equation}\label{e20}
\varphi_{a}   \rightarrow   \lambda_{ab} \varphi_{b}
\end{equation}
which means
\begin{equation}\label{e21}
\Phi \rightarrow det( \lambda_{ab}) \Phi \\ \equiv \lambda
\Phi     \end{equation}
such that
\begin{equation}\label{e22}
\lambda = e^{\theta}
\end{equation}
and
\begin{equation}\label{e23}
\phi \rightarrow \phi - \frac{\theta}{\alpha}.
\end{equation}

We will now work out the equations of motion after introducing $V(\phi)$ and $U(\phi)$
and see how the integration of the equations of motion allows the spontaneous breaking of the
scale invariance.  

    Let us begin by considering the equations which are obtained from
the variation of the fields that appear in the measure, i.e. the
$\varphi_{a}$
fields. We obtain then
\begin{equation}\label{e24}
A^{\mu}_{a} \partial_{\mu} L_{2} = 0
\end{equation}
where  $A^{\mu}_{a} = \varepsilon^{\mu\nu\alpha\beta}
\varepsilon_{abcd} \partial_{\nu} \varphi_{b} \partial_{\alpha}
\varphi_{c} \partial_{\beta} \varphi_{d}$. Since it is easy to
check that  $A^{\mu}_{a} \partial_{\mu} \varphi_{a^{\prime}} =
\frac{\delta aa^{\prime}}{4} \Phi$, it follows that
det $(A^{\mu}_{a}) =\frac{4^{-4}}{4!} \Phi^{3} \neq 0$ if $\Phi\neq 0$.
Therefore if $\Phi\neq 0$ we obtain that $\partial_{\mu} L_{2} = 0$,
 or that
\begin{equation}\label{e25}
L_{2} = \frac{-1}{\kappa} R(\Gamma,g) + \frac{1}{2} g^{\mu\nu}
\partial_{\mu} \phi \partial_{\nu} \phi - V = M
\end{equation}
where M is constant. Notice that this equation breaks spontaneously the global scale invariance of the theory, 
since the left hand side has a non trivial transformation under the scale transformations, while the right
hand side is equal to $M$, a constant that after we integrate the equations is fixed, cannot be changed and therefore
for any $M\neq 0$ we have obtained indeed, spontaneous breaking of scale invariance.

We will see what is the connection now. As we will see, the connection appears in the original frame as a
non Riemannian object. However, we will see that by a simple conformal tranformation of the metric we can recover
the Riemannian structure. The interpretation of the equations in the frame gives then an interesting physical picture, 
as we will see.

    Let us begin by studying the equations obtained from the variation of the
connections $\Gamma^{\lambda}_{\mu\nu}$.  We obtain then
\begin{equation}\label{e26}
-\Gamma^{\lambda}_{\mu\nu} -\Gamma^{\alpha}_{\beta\mu}
g^{\beta\lambda} g_{\alpha\nu}  + \delta^{\lambda}_{\nu}
\Gamma^{\alpha}_{\mu\alpha} + \delta^{\lambda}_{\mu}
g^{\alpha\beta} \Gamma^{\gamma}_{\alpha\beta}
g_{\gamma\nu}\\ - g_{\alpha\nu} \partial_{\mu} g^{\alpha\lambda}
+ \delta^{\lambda}_{\mu} g_{\alpha\nu} \partial_{\beta}
g^{\alpha\beta}
 - \delta^{\lambda}_{\nu} \frac{\Phi,_\mu}{\Phi}
+ \delta^{\lambda}_{\mu} \frac{\Phi,_           \nu}{\Phi} =  0
\end{equation}
If we define $\Sigma^{\lambda}_{\mu\nu}$    as
$\Sigma^{\lambda}_{\mu\nu} =
\Gamma^{\lambda}_{\mu\nu} -\{^{\lambda}_{\mu\nu}\}$
where $\{^{\lambda}_{\mu\nu}\}$   is the Christoffel symbol, we
obtain for $\Sigma^{\lambda}_{\mu\nu}$ the equation
\begin{equation}\label{e27}
    -  \sigma, _{\lambda} g_{\mu\nu} + \sigma, _{\mu}
g_{\nu\lambda} - g_{\nu\alpha} \Sigma^{\alpha}_{\lambda\mu}
-g_{\mu\alpha} \Sigma^{\alpha}_{\nu \lambda}
+ g_{\mu\nu} \Sigma^{\alpha}_{\lambda\alpha} +
g_{\nu\lambda} g_{\alpha\mu} g^{\beta\gamma} \Sigma^{\alpha}_{\beta\gamma}
= 0
\end{equation}
where  $\sigma = ln \chi, \chi = \frac{\Phi}{\sqrt{-g}}$.

    The general solution of eq.(\ref{e28}) is
\begin{equation}\label{e28}
\Sigma^{\alpha}_{\mu\nu} = \delta^{\alpha}_{\mu}
\lambda,_{\nu} + \frac{1}{2} (\sigma,_{\mu} \delta^{\alpha}_{\nu} -
\sigma,_{\beta} g_{\mu\nu} g^{\alpha\beta})
\end{equation}\label{e30}
where $\lambda$ is an arbitrary function due to the $\lambda$ - symmetry
of the
curvature \cite{Lambda} $R^{\lambda}_{\mu\nu\alpha} (\Gamma)$,
\begin{equation}\label{e29}
\Gamma^{\alpha}_{\mu\nu} \rightarrow \Gamma^{\prime \alpha}_{\mu\nu}
 = \Gamma^{\alpha}_{\mu\nu} + \delta^{\alpha}_{\mu}
Z,_{\nu}
\end{equation}
Z  being any scalar (which means $\lambda \rightarrow \lambda + Z$).

    If we choose the gauge $\lambda = \frac{\sigma}{2}$, we obtain
\begin{equation}\label{e30}
\Sigma^{\alpha}_{\mu\nu} (\sigma) = \frac{1}{2} (\delta^{\alpha}_{\mu}
\sigma,_{\nu} +
 \delta^{\alpha}_{\nu} \sigma,_{\mu} - \sigma,_{\beta}
g_{\mu\nu} g^{\alpha\beta}).
\end{equation}

    Considering now the variation with respect to $g^{\mu\nu}$, we
obtain
\begin{equation}\label{e31}
\Phi (\frac{-1}{\kappa} R_{\mu\nu} (\Gamma) + \frac{1}{2} \phi,_{\mu}
\phi,_{\nu}) - \frac{1}{2} \sqrt{-g} U(\phi) g_{\mu\nu} = 0
\end{equation}
solving for $R = g^{\mu\nu} R_{\mu\nu} (\Gamma)$  from eq.(\ref{e31}) and introducing
in eq.(\ref{e25}), we obtain
\begin{equation}\label{e32}
M + V(\phi) - \frac{2U(\phi)}{\chi} = 0
\end{equation}
a constraint that allows us to solve for $\chi$,
\begin{equation}\label{e33}
\chi = \frac{2U(\phi)}{M+V(\phi)}.
\end{equation}

    To get the physical content of the theory, it is best
consider variables that have well defined dynamical interpretation. The original
metric does not has a non zero canonical  momenta. The fundamental
variable of the theory in the first order formalism is the connection and its
canonical momenta is a function of $\overline{g}_{\mu\nu}$, given by,

\begin{equation}\label{e34}
\overline{g}_{\mu\nu} = \chi g_{\mu\nu}
\end{equation}

and $\chi$  given by eq.(\ref{e33}). Interestingly enough, working with $\overline{g}_{\mu\nu}$
is the same as going to the "Einstein Conformal Frame".
In terms of $\overline{g}_{\mu\nu}$   the non
Riemannian contribution $\Sigma^{\alpha}_{\mu\nu}$
dissappears from the equations. This is because the connection
can be written as the Christoffel symbol of the metric
$\overline{g}_{\mu\nu}$ .
In terms of $\overline{g}_{\mu\nu}$ the equations
of motion for the metric can be written then in the Einstein
form (we define $\overline{R}_{\mu\nu} (\overline{g}_{\alpha\beta}) =$
 usual Ricci tensor in terms of the bar metric $= R_{\mu\nu}$ and
 $\overline{R}  = \overline{g}^{\mu \nu}  \overline{R}_{\mu\nu}$ )
\begin{equation}\label{e35}
\overline{R}_{\mu\nu} (\overline{g}_{\alpha\beta}) - \frac{1}{2}
\overline{g}_{\mu\nu}
\overline{R}(\overline{g}_{\alpha\beta}) = \frac{\kappa}{2} T^{eff}_{\mu\nu}
(\phi)
\end{equation}
where
\begin{equation}\label{e36}
T^{eff}_{\mu\nu} (\phi) = \phi_{,\mu} \phi_{,\nu} - \frac{1}{2} \overline
{g}_{\mu\nu} \phi_{,\alpha} \phi_{,\beta} \overline{g}^{\alpha\beta}
+ \overline{g}_{\mu\nu} V_{eff} (\phi)
\end{equation}

and
\begin{equation}\label{e37}
V_{eff} (\phi) = \frac{1}{4U(\phi)}  (V+M)^{2}.
\end{equation}

    In terms of the metric $\overline{g}^{\alpha\beta}$ , the equation
of motion of the Scalar
field $\phi$ takes the standard General - Relativity form
\begin{equation}\label{e38}
\frac{1}{\sqrt{-\overline{g}}} \partial_{\mu} (\overline{g}^{\mu\nu}
\sqrt{-\overline{g}} \partial_{\nu}
\phi) + V^{\prime}_{eff} (\phi) = 0.
\end{equation}

    Notice that if  $V + M = 0,  V_{eff}  = 0$ and $V^{\prime}_{eff}
= 0$ also, provided $V^{\prime}$ is finite and $U \neq 0$ there.
This means the zero cosmological constant
state
is achieved without any sort of fine tuning. That is, independently
of whether we add to $V$ a constant piece, or whether we change
the value of $M$, as long as there is still a point
where $V+M =0$, then still $ V_{eff}  = 0$ and $V^{\prime}_{eff} = 0$
( still provided $V^{\prime}$ is finite and $U \neq 0$ there).
This is the basic feature
that characterizes the TMT and allows it to solve the 'old'
cosmological constant problem, at least at the classical level.

    In what follows we will study the effective potential (\ref{e37}) for the special case of global
scale invariance, which as we will see displays additional very special
features which makes it attractive in the context of cosmology.

    Notice that in terms of the variables $\phi$,
$\overline{g}_{\mu\nu}$, the "scale"
transformation becomes only a shift in the scalar field $\phi$, since
$\overline{g}_{\mu\nu}$ is
invariant (since $\chi \rightarrow \lambda^{-1} \chi$  and $g_{\mu\nu}
\rightarrow \lambda g_{\mu\nu}$)
\begin{equation}\label{e39}
\overline{g}_{\mu\nu} \rightarrow \overline{g}_{\mu\nu}, \phi \rightarrow
\phi - \frac{\theta}{\alpha}.
\end{equation}

    If $V(\phi) = f_{1} e^{\alpha\phi}$  and  $U(\phi) = f_{2}
e^{2\alpha\phi}$ as
required by scale
invariance eqs.(\ref{e18}),(\ref{e20}),(\ref{e21}),(\ref{e22}),(\ref{e23}), we obtain from the expression (\ref{e37})
\begin{equation}\label{e40}
    V_{eff}  = \frac{1}{4f_{2}}  (f_{1}  +  M e^{-\alpha\phi})^{2}
\end{equation}

    Since we can always perform the transformation $\phi \rightarrow
- \phi$ we can
choose by convention $\alpha > 0$. We then see that as $\phi \rightarrow
\infty, V_{eff} \rightarrow \frac{f_{1}^{2}}{4f_{2}} =$ const.
providing an infinite flat region. Also a minimum is achieved at zero
cosmological constant for the case $\frac{f_{1}}{M} < 0$ at the point
\begin{equation}\label{e41}
\phi_{min}  =  \frac{-1}{\alpha} ln \mid\frac{f_1}{M}\mid.
\end{equation}

    In conclusion, the scale invariance of the original theory is
responsible for the non appearance (in the physics) of a certain scale,
that associated to M. However, masses do appear, since the coupling to two
different measures of $L_{1}$ and $L_{2}$ allow us to introduce two
independent
couplings  $f_{1}$ and $f_{2}$, a situation which is  unlike the
standard
formulation of globally scale invariant theories, where usually no stable
vacuum state exists.

The constant of integration $M$ plays a very important role indeed:
any non vanishing value for this constant implements, already at the
classical level S.S.B. of scale invariance.

\section{ Generation of two flat regions after the introduction of a $R^{2}$ term}

As we have seen, it is possible to obtain a model that through a spontaneous breaking of scale invariace
can give us a flat region. We want to obtain now two flat regions in our effective potential.
A simple generalization of the action $S_{L}$ will fix this.
What one needs to do is simply consider  the
addition of a scale invariant term of the form

\begin{equation}\label{e45}
S_{R^{2}} = \epsilon  \int (g^{\mu\nu} R_{\mu\nu} (\Gamma))^{2} \sqrt{-g} d^{4} x
\end{equation}

The total action being then $S = S_{L} + S_{R^{2}}$.
In the first order formalism $ S_{R^{2}}$ is not only globally scale invariant
but also locally scale invariant, that is conformally invariant (recall that
in the first order formalism the connection is an independent degree of freedom
and it does not transform under a conformal transformation of the metric). The higher curvature theories
in the context of the second order formalism \cite{barrow1}-\cite{barrow6},\cite{mijic1}- \cite{mijic3} have a completly different behavior, giving higher order equations, etc., unlike higher curvature theories in the context of the first order formalism, like we do here.

Let us see what the equations of motion tell us, now with the addition of
$S_{R^{2}}$ to the action. First of all, since the addition has been only to
the part of the action that couples to $ \sqrt{-g}$, the equations of motion
derived from the variation of the measure fields remains unchanged. That is
eq.(\ref{e25}) remains valid.

The variation of the action with respect to $ g^{\mu \nu}$ gives now

\begin{equation}\label{e46}
 R_{\mu\nu} (\Gamma) ( \frac{-\Phi}{\kappa} + 2 \epsilon R  \sqrt{-g}) +
\Phi \frac{1}{2} \phi,_{\mu} \phi,_{\nu} -
\frac{1}{2}(\epsilon R^{2} + U(\phi)) \sqrt{-g} g_{\mu\nu} = 0
\end{equation}

It is interesting to notice that if we contract this equation with
 $ g^{\mu \nu}$ , the $\epsilon$ terms do not contribute. This means
that the same value for the scalar curvature $R$ is obtained as in section II,
 if we express our result in terms of $\phi$, its derivatives and
$ g^{\mu \nu}$ .
Solving the scalar curvature from this and inserting in the other
$\epsilon$ - independent equation $L_{2} = M$  we get still the same
solution for the ratio of the measures which was found in the case where
the $\epsilon$ terms were absent,
i.e. $\chi =  \frac{\Phi}{\sqrt{-g}}  = \frac{2U(\phi)}{M+V(\phi)}$.

In the presence of the $\epsilon R^{2} $ term in the action, eq. (\ref{e26})
gets modified so that instead of $\Phi$,  $\Omega$  =
$\Phi - 2 \epsilon R \sqrt{-g}$ appears. This in turn implies that
eq.(\ref{e27}) mantains its form but where $\sigma$ is replaced by
$\omega  = ln (\frac{\Omega}{\sqrt{-g}}) =
 ln ( \chi -2\kappa \epsilon R)$,
where once again,
$\chi =  \frac{\Phi}{\sqrt{-g}} = \frac{2U(\phi)}{M+V(\phi)}$.

Following then the same steps as in the model without the curvature square terms, we can then verify that the
connection is the Christoffel symbol of the metric $\overline{g}_{\mu\nu}$
given by

\begin{equation}\label{e47}
\overline{g}_{\mu\nu}   = (\frac{\Omega}{\sqrt{-g}}) g_{\mu\nu}
 = (\chi -2\kappa \epsilon R) g_{\mu\nu}
\end{equation}

$\overline{g}_{\mu\nu} $ defines now the "Einstein frame". Equations (\ref{e46})
can now be expressed in the "Einstein form"

\begin{equation}\label{e48}
\overline{R}_{\mu\nu} -  \frac{1}{2}\overline{g}_{\mu\ \nu} \overline{R} =
\frac{\kappa}{2} T^{eff}_{\mu\nu}
\end{equation}

where

\begin{equation}
 T^{eff}_{\mu\nu} =\label{e49}
\frac{\chi}{\chi -2 \kappa \epsilon R} (\phi_{,\mu} \phi_{,\nu} - \frac{1}{2} \overline
{g}_{\mu\nu} \phi_{,\alpha} \phi_{,\beta} \overline{g}^{\alpha\beta})
+ \overline{g}_{\mu\nu} V_{eff}
\end{equation}

where

\begin{equation}\label{e50}
 V_{eff}  = \frac{\epsilon R^{2} + U}{(\chi -2 \kappa \epsilon R)^{2} }
\end{equation}

Here it is satisfied that $\frac{-1}{\kappa} R(\Gamma,g) +
\frac{1}{2} g^{\mu\nu}\partial_{\mu} \phi \partial_{\nu} \phi - V = M $,
equation that expressed in terms of $ \overline{g}^{\alpha\beta}$
 becomes

$\frac{-1}{\kappa} R(\Gamma,g) + (\chi -2\kappa \epsilon R)
\frac{1}{2} \overline{g}^{\mu\nu}\partial_{\mu} \phi \partial_{\nu} \phi - V = M$.
 This allows us to solve for $R$ and we get,

\begin{equation}\label{e51}
R = \frac{-\kappa (V+M) +\frac{\kappa}{2} \overline{g}^{\mu\nu}\partial_{\mu} \phi \partial_{\nu} \phi \chi}
{1 + \kappa ^{2} \epsilon \overline{g}^{\mu\nu}\partial_{\mu} \phi \partial_{\nu} \phi}
\end{equation}

Notice that
 if we express $R$ in
terms of $\phi$, its derivatives and $ g^{\mu \nu}$, the result is the
same as in the model without the curvature squared term, this is not true anymore once we express
 $R$ in terms of $\phi$, its derivatives and $\overline{g}^{\mu\nu}$.

In any case, once we insert (\ref{e51}) into (\ref{e50}),
we see that the  effective potential  \ref{e50} will depend on the derivatives of the
scalar field now. It acts as a normal scalar field potential under the
conditions of slow rolling  or low gradients and in the case the
scalar field is near the region $M+V(\phi ) = 0$.

Notice that since
$\chi =   \frac{2U(\phi )}{M+V(\phi )}$,
then if ${M+V(\phi) = 0}$, then, as in the simpler model without the curvature squared terms, we obtain that
 $ V_{eff}  =  V'_{eff} =  0$ at that point without fine tuning
(here by $ V'_{eff}$ we mean the derivative  of $ V_{eff}$ with
respect to the scalar field $\phi$, as usual).

In the case of the scale invariant case, where $V$ and $U$ are given by
equation (\ref{e19}), it is interesting to study the shape of $ V_{eff} $
as a function of $\phi$
in the case of a constant $\phi$, in which case $ V_{eff} $ can be
regarded as a real scalar field potential. Then from (\ref{e51}) we get
$R = -\kappa (V+M)$, which inserted in (\ref{e50}) gives,
\begin{equation}\label{effpotslow}
 V_{eff}  =
\frac{(f_{1} e^{ \alpha \phi }  +  M )^{2}}{4(\epsilon \kappa ^{2}(f_{1}e^{\alpha \phi}  +  M )^{2} + f_{2}e^{2 \alpha \phi })}
\end{equation}

The limiting values of $ V_{eff} $ are:

First, for asymptotically
large positive values, ie. as $ \alpha\phi \rightarrow  \infty $,
we have
$V_{eff} \rightarrow
\frac{f_{1}^{2}}{4(\epsilon \kappa ^{2} f_{1}^{2} + f_{2})} $.

Second, for asymptotically large but negative values of the scalar field,
that is as $\alpha \phi \rightarrow - \infty  $ ,  we have:
$ V_{eff} \rightarrow \frac{1}{4\epsilon \kappa ^{2}}$ .

In these two asymptotic regions ($\alpha \phi \rightarrow  \infty  $
and $\alpha \phi \rightarrow - \infty  $) an examination of the scalar
field equation reveals that a constant scalar field configuration is a
solution of the equations, as is of course expected from the flatness of
the effective potential in these regions.

Notice that in all the above discussion it is fundamental that $ M\neq 0$.
If $M = 0$ the potential becomes just a flat one,
$V_{eff} = \frac{f_{1}^{2}}{4(\epsilon \kappa ^{2} f_{1}^{2} + f_{2})}$
everywhere (not only at high values  of $\alpha \phi$). All the non trivial
features necessary for a gracefull exit, the other flat
region associated to the Planck scale and the minimum at zero if $M<0$ are all lost .
As we discussed in the model without a curvature squared term, $ M\neq 0$ implies the we are considering a
situation with S.S.B. of scale invariance.

These kind of models with potentials giving rise to two flat potentials have been applied to produce models for bags and confinement in a very natural way \cite{bags and confinement}

\section{A Note on the the "Einstein" metric as a canonical variable of the Theory}
One could question the use of the Einstein frame metric $\overline{g}_{\mu\nu} $ in contrast to the original metric $g_{\mu\nu} $ . In this respect, 
it is interesting to see the role of both the original metric and that of the Einstein frame metric
in a canonical approach to the first order formalism. Here we see that the original metric does not have a canonically conjugated momentum (this turns out to be zero), in contrast, the canonically conjugated momentum to the conection turns out to be a function exclusively of 
$\overline{g}_{\mu\nu} $, this Einstein metric is therefore a genuine dynamical canonical variable, as opposed to the original metric. There is also a lagrangian formulation of the theory which uses $\overline{g}_{\mu\nu} $, as we will see in the next section, what we can call the action in the Einstein frame. In this frame we can quantize the theory for example and consider contributions without reference to the original frame, thus possibly considering breking the TMT structure of the theory through quantum effects, but such breaking will be done "softly" through the introduction of a cosmological term only. Surpringly, the remaining structure of the theory, reminiscent from the original TMT structure will be enough
to control the strength of this additional cosmological term once we demand that the universe originated from a non singular and stable emergent state.

\section{Generalizing the model to include effects of zero point fluctuations}

The effective energy-momentum tensor can be
represented in a form like that of  a perfect fluid
\begin{equation}
T_{\mu\nu}^{eff}=(\rho +p)u_{\mu}u_{\nu}-p\bar{g}_{\mu\nu},
\qquad \text{where} \qquad
u_{\mu}=\frac{\phi_{,\mu}}{(2X)^{1/2}}\label{Tmnfluid}
\end{equation}
here $X\equiv\frac{1}{2}\bar{g}^{\alpha\beta}\phi_{,\alpha}\phi_{,\beta}$. This defines a pressure functional and an energy density functional.
The system of equations obtained after solving for $\chi$, working 
in the Einstein frame with the metric
$\bar{g}_{\mu \nu}$ can be obtained from a
 "k-essence" type effective action, as it is standard in treatments 
 of theories with non linear kinetic tems or k-essence models\cite{k-essence1}-\cite{k-essence4}. The action from which the classical equations follow is, 
\begin{equation}
S_{eff}=\int\sqrt{-\bar{g}}d^{4}x\left[-\frac{1}{\kappa}\bar{R}(\bar{g})
+p\left(\phi,R\right)\right] \label{k-eff}
\end{equation}

\begin{equation}
 p = \frac{\chi}{\chi -2 \kappa \epsilon R}X - V_{eff}
\end{equation}

\begin{equation}\label{e500}
 V_{eff}  = \frac{\epsilon R^{2} + U}{(\chi -2 \kappa \epsilon R)^{2} }
\end{equation}

where it is understood that,
\begin{equation}\label{chi}
\chi = \frac{2U(\phi)}{M+V(\phi)}.
\end{equation}
We have two possible formulations concerning $R$:
Notice first that $\bar{R}$ and $R$ are different objects, the $\bar{R}$ is the Riemannian curvature scalar in the Einstein frame,
while $R$ is a different object. This $R$ will be treated in two different ways:

1. First order formalism for $R$. Here $R$ is a lagrangian variable, determined as follows,  $R$ that appear in the expression above for $p$ can be obtained from the variation of the pressure functional action above with respect to $R$, this gives exactly the expression for $R$ that has been solved already in terms of $X, \phi$, etc. 

2. Second order formalism for $R$. $R$ that appear in the action above is exactly the expression for $R$ that has been solved already in terms of $X, \phi$, etc. The second order formalism can be obtained from the first order formalism by solving algebraically R from the eq. obtained by variation of $R$ , and inserting back into the action.

In contrast to the
simplified models studied in literature\cite{k-essence1}-\cite{k-essence4}, it is
impossible here to represent $p\left(\phi,X;M\right)$ in a
factorizable form like $\tilde{K}(\phi)\tilde{p}(X)$. The scalar
field effective Lagrangian can be taken as a starting point  
for many considerations.

In particular, the quantization of the model can proceed from (\ref{k-eff}) and additional terms could
be generated by radiative corrections. We will focus only on a possible cosmological term in
the Einstein frame added (due to zero point fluctuations) to (\ref{k-eff}), which leads then to the new action
\begin{equation}
S_{eff,\Lambda }=\int\sqrt{-\bar{g}}d^{4}x\left[-\frac{1}{\kappa}\bar{R}(\bar{g})
+p\left(\phi,R\right)- \Lambda \right] \label{act.lambda}
\end{equation}

This addition to the effective action leaves the equations of motion of the scalar field unaffected, but the gravitational equations aquire a 
cosmological constant. Adding the $\Lambda$ term can be regarded as a redefinition of $V_{eff}\left(\phi,X;M\right)$
\begin{equation}
V_{eff}\left(\phi,R\right) \rightarrow V_{eff}\left(\phi,R\right) + \Lambda  \label{V.lambda}
\end{equation}As we will see the stability of the emerging Universe imposses interesting constraints on $\Lambda$

After introducing the $\Lambda$ term, we get from the variation of $R$ the same value of $R$, unaffected by the new $\Lambda$ term, but as one can easily see then $R$ does not have the interpretation of a curvature scalar in the original frame since it is unaffected by the new source of energy density (the $\Lambda$ term), this is why the $\Lambda$ term theory does not have a formulation in the original frame, but is a perfectly legitimate generalization of the theory, probably obtained by considering zero point fluctuations, notice that quantum theory is possible only in the Einstein frame. Notice that even in the original frame  the bar metric (not the original metric) appears automatically in the canonically conjugate momenta to the connection, so we can expect from this that the bar metric and not the original metric be the relevant one for the quantum theory.

\section{Analysis of the Emergent Universe solutions}
We now want to consider the detailed analysis of The Emerging Universe solutions and in the next section their stability
in the TMT scale invariant theory.
We start considering the cosmological solutions of the form (in the Einstein frame),
\begin{equation}
ds^2 =dt^2 - a(t)^2 (\frac{dr^2}{1 -r^2}+ r^2(d\theta^2 +sin^2\theta d\phi^2)),   \phi = \phi(t)
\end{equation}

in this case, we obtain for the energy density and the pressure, the following expressions.
We will consider a scenario where the scalar field $\phi$ is moving in the extreme right region
$\phi \rightarrow \infty  $, in this case the expressions for the energy density $\rho$ and pressure $p$ are given by,
\begin{equation}\label{eq.density}
\rho = \frac{A}{2} \dot{\phi}^2 + 3B\dot{\phi}^4 + C
\end{equation}

and
\begin{equation}
p = \frac{A}{2} \dot{\phi}^2 +B\dot{\phi}^4 - C
\end{equation}
It is interesting to notice that all terms proportional to $\dot{\phi}^4$ behave like "radiation", since
 $p_{\dot{\phi}^4} = \frac{\rho_{\dot{\phi}^4} }{3}$ is satisfied. 
here the constants $A,B$ and $C$ are given by,

\begin{eqnarray}\label{ABC}
A &=& \frac{f_2}{f_2 + \kappa^2\epsilon f_1^2}\,,\\
B &=& \frac{\epsilon\kappa^2}{4(1+\kappa^2\epsilon f_1^2/f_2)} = \frac{\epsilon \kappa^2}{4}\,A \,,\label{B} \\
C &=& \frac{f_1^2}{4\,f_2(1+\kappa^2\epsilon f_1^2/f_2)} =
\frac{f_1^2}{4f_2}\,A + \Lambda\,\label{C}.
\end{eqnarray}
It will be convenient to "decompose" the constant $\Lambda$ into two pieces, 

\begin{eqnarray}\label{Lambda}
\Lambda = -\frac{1}{4\kappa^2\epsilon} + \Delta \lambda
\end{eqnarray}
since as $\phi \rightarrow -\infty $ , $ V_{eff} \rightarrow  \Delta \lambda $. Therefore $\Delta \lambda$ has the interesting interpretation of the vacuum energy density in the $\phi \rightarrow -\infty $ vacuum. As we will see, it is remarkable that the stability and existence of non singular emergent universe implies that $\Delta \lambda > 0$, and it is bounded from above as well.

The equation that determines such static universe $a(t) = a_0 =constant$,
$\dot{a}=0$, $\ddot{a}=0$ gives rise to a restriction for $\dot{\phi}_0$
that have to satisfy the following equation in order to
guarantee that the universe be static, because $\ddot{a}=0$ is proportional to 
$\rho + 3p$, we must require that $\rho + 3p = 0$, which leads to

\begin{equation}\label{e.1}
3B\dot{\phi}^4_0 + A\dot{\phi}^2_0 - C=0,
\end{equation}

This equation leads to two roots, the first being

\begin{equation}\label{e.2}
\dot{\phi}_1^2=\frac{\sqrt{A^2+ 12BC}\,-A}{6B}\,.
\end{equation}

The second root is:

\begin{equation}\label{e.3}
\dot{\phi}_2^2=\frac{-\sqrt{A^2+ 12BC}\,-A}{6B}\,.
\end{equation}

It is also interesting to see that if the discriminant is positive, the first solution has automatically positive energy density, if we only consider cases where $C>0$, which is required if we want the emerging solution to be able to turn into an inflationary solution eventually. One can see that the condition $\rho >0$ for the first solution reduces to the inequality $w> (1-\sqrt{1-w} )/2$, where $w =-12BC/A^2 >0$, since we must have $A>0$, otherwise we get a negative kinetic term during the inflationary period, and as we will see in the next section, we must have that $B<0$ from the stability of the solution, and as long as the discriminant is positive, i.e. $0<w<1$, it is always true that this inequality is satisfied.

\section{Stability of the static solution}

We will now consider the
perturbation equations. Considering small deviations of $\dot{\phi}$ the from the static emerging solution
value $\dot{\phi}_0$ and also considering the perturbations of the scale factor $a$, we obtain, from
Eq.~(\ref{eq.density})

\begin{equation}\label{eq.density-pert.}
\delta \rho = A \dot{\phi}_0 \delta \dot{\phi} + 12B \dot{\phi}_0^3 \delta \dot{\phi}
\end{equation}

at the same time $\delta \rho$ can be obtained from the perturbation of the Friedmann equation

\begin{equation}\label{Fried.eq.}
3(\frac{1}{a^2}+H^2)=\kappa \rho 
\end{equation}
and since we are perturbing a solution which is static, i.e., has $H=0$, we obtain then
\begin{equation}\label{pert.Fried.eq.}
-\frac{6}{a_0^3}\delta a =\kappa \delta \rho 
\end{equation}

we also have the second order Friedmann equation

\begin{equation}\label{Fried.eq.2}
\frac{1+\dot{a}^2 + 2a\ddot{a}}{a^2}=-\kappa p 
\end{equation}

For the static emerging solution, we have $p_0=-\rho_0/3$, $a=a_0$,  so 
\begin{equation}
\frac{2}{a_0^2} = -2\kappa p_0 = \frac{2}{3}\kappa \rho_0= \Omega_0 \kappa \rho_0
\end{equation}
where we have chosen to express our result in terms of $\Omega_0$, defined by $p_0=(\Omega_0-1)\rho_0$, which for the emerging
solution has the value $\Omega_0=\frac{2}{3}$. Using this in (\ref{pert.Fried.eq.}), we obtain
\begin{equation}\label{pert.Fried.eq.3}
\delta \rho = -\frac{3\Omega_0 \rho_0}{a_0}\delta a  
\end{equation}
and equating the values of $\delta \rho$ as given by (\ref{eq.density-pert.}) and (\ref{pert.Fried.eq.3}) we obtain a linear relation between
$\delta \dot{\phi}$ and $\delta a$, which is, 
\begin{equation}\label{delta-delta}
\delta \dot{\phi}=D_0\delta a 
\end{equation}
where 

\begin{equation}  
D_0 = -\frac{3\Omega_0 \rho_0}{a_0 \dot{\phi}_0 (A + 12 B \dot{\phi}_0^2)}
\end{equation}

we now consider the perturbation of the eq. (\ref{Fried.eq.2}). In the right hand side of this equation we consider
that $p=(\Omega-1)\rho$, with
\begin{equation}\label{Omega-eq.}
\Omega = 2\Big(1 - \frac{U_{eff}}{\rho}\Big),
\end{equation}
where,
\begin{equation}\label{V-eq.}
U_{eff} =C + B\,\dot{\phi}^4
\end{equation}

and therefore,  the perturbation of the eq. (\ref{Fried.eq.2}) leads to,

\begin{equation}\label{pert.Fried.eq.2}
-\frac{2\delta a}{a_0^3}+2\frac{\delta\ddot{a}}{a_0}=-\kappa \delta p =-\kappa \delta ((\Omega-1)\rho)
\end{equation}

to evaluate this, we use (\ref{Omega-eq.}), (\ref{V-eq.}) and the expressions that relate the variations in $a$
and $\dot{\phi}$ (\ref{delta-delta}).  Defining the "small"  variable $\beta$ as

\begin{equation}
a(t) = a_0( 1+ \beta)
\end{equation}
we obtain,
\begin{equation}
2\ddot{\beta}(t) + W_0^2\beta(t) = 0\,,
\end{equation}

where,
\begin{equation}
W_0^2 = \Omega_0\,\rho_0\left[ \frac{24\,B\,\dot{\phi}_0^2}{A +
12\,\dot{\phi}_0^2\,B }  -6\frac{(C + B\,
\dot{\phi}_0^4)}{\rho_0} -3\kappa \Omega_0 + 2\kappa \right],
\end{equation}

notice that the sum of the last two terms in the expression for $W_0^2$, that is $-3\kappa \Omega_0 + 2\kappa $  vanish
since $\Omega_0=\frac{2}{3}$, for the same reason, we have that $6\frac{(C + B\,\dot{\phi}_0^4)}{\rho_0} = 4$, which brings us to the simplified expression
\begin{equation}
W_0^2 = \Omega_0\,\rho_0\left[ \frac{24\,B\,\dot{\phi}_0^2}{A +
12\,\dot{\phi}_0^2\,B }  - 4 \right],
\end{equation}

For the stability of the static solution, we need that  $W_0^2 >0$,
where $\dot{\phi}_0^2$ is defined either by E.~(\ref{e.2}) ($\dot{\phi}_0^2=\dot{\phi}_1^2$) or by E.~(\ref{e.3}) 
($\dot{\phi}_0^2=\dot{\phi}_2^2$). 
If we take E.~(\ref{e.3}) ($\dot{\phi}_0^2=\dot{\phi}_2^2$) and use this in the above expression for $W_0^2$, we obtain,
\begin{equation}
W_0^2 = \Omega_0\,\rho_0\left[ \frac{4\sqrt{A^2 +12BC}}{-2\sqrt{A^2 +12BC} -A}  \right],
\end{equation}

to avoid negative kinetic terms during the slow roll phase that takes place following the emergent phase, we must consider $A>0$, so, we  see that the second solution is unstable and will not be considered further. 

Now in the case of the first solution, E.~(\ref{e.2}) ($\dot{\phi}_0^2=\dot{\phi}_1^2$), then $W_0^2$ becomes
\begin{equation}
W_0^2 = \Omega_0\,\rho_0\left[ \frac{-4\sqrt{A^2 +12BC}}{2\sqrt{A^2 +12BC} -A}  \right],
\end{equation}
so the condition of stability becomes $2\sqrt{A^2 +12BC} -A < 0$, or $2\sqrt{A^2 +12BC} < A  $, squaring both sides and since $A>0$, we get
$12BC/A^2 < -3/4$, which means $B<0$, and therefore $\epsilon <0$,  multiplying by $-1$, we obtain, $12(-B)C/A^2 > 3/4$, replacing the values of $A, B, C$, given by (\ref{ABC}) we obtain the condition

\begin{equation}
 \Delta\lambda > 0,
\end{equation}

Now there is the condition that the discriminant be positive $A^2 +12BC> 0$

\begin{equation}
 \Delta\lambda < \frac{1}{12(-\epsilon)\kappa^2 } \left[ \frac{f_2}{f_2 + \kappa^2 \epsilon f^2_1 }  \right],
\end{equation}

since $A=\left[ \frac{f_2}{f_2 + \kappa^2 \epsilon f^2_1 }  \right]> 0$, $B<0$, meaning that $\epsilon < 0$, we see that we obtain a positive upper bound for the energy density of the vacuum as $\phi \rightarrow -\infty $, which must be positive, but not very big.
 \section{The vacuum structure of the theory. Evolution of the Universe, from its non singular origins to its present slowly accelerating state at $\phi \rightarrow -\infty$ , crossing "barriers".}
 For the discussion of the vacuum structure of the theory, we  start studying $ V_{eff} $ for the case of a constant field  $\phi$, given by,
 \begin{equation}\label{effppluslambda}
 V_{eff}  =
\frac{(f_{1} e^{ \alpha \phi }  +  M )^{2}}{4(\epsilon \kappa ^{2}(f_{1}e^{\alpha \phi}  +  M )^{2} + f_{2}e^{2 \alpha \phi })}+ \Lambda
\end{equation}

 This is necessary, but not enough, since as we will see, the consideration of  constant fields  $\phi$ alone can lead to missleading conclusions, in some cases, the dependence of $ V_{eff} $ on the kinetic term can be crucial to see if and how we can achieve the crossing of an apparent barrier.

For a constant field  $\phi$ the limiting values of $ V_{eff} $ are (now that we added the constant $\Lambda $):

First, for asymptotically
large positive values, ie. as $ \alpha\phi \rightarrow  \infty $,
we have
$V_{eff} \rightarrow
\frac{f_{1}^{2}}{4(\epsilon \kappa ^{2} f_{1}^{2} + f_{2})}+\Lambda $.

Second, for asymptotically large but negative values of the scalar field,
that is as $\alpha \phi \rightarrow - \infty  $ ,  we have:
$ V_{eff} \rightarrow \frac{1}{4\epsilon \kappa ^{2}}+\Lambda = \Delta \lambda $ .

In these two asymptotic regions ($\alpha \phi \rightarrow  \infty  $
and $\alpha \phi \rightarrow - \infty  $) an examination of the scalar
field equation reveals that a constant scalar field configuration is a
solution of the equations, as is of course expected from the flatness of
the effective potential in these regions.

Notice that in all the above discussion it is fundamental that $ M\neq 0$.
If $M = 0$ the potential becomes just a flat one,
$V_{eff} = \frac{f_{1}^{2}}{4(\epsilon \kappa ^{2} f_{1}^{2} + f_{2})}+\Lambda $
everywhere (not only at high values  of $\alpha \phi$).

Finally, there is a minimum at $V_{eff}= \Lambda $ if $M<0$ .
In summary, and if $f_2>0$, $A>0$, we have that there is a hierarchy of vacua ,

\begin{equation}
V_{eff} (\alpha \phi \rightarrow - \infty )=  \Delta \lambda< V_{eff}(min, M < 0)= \Lambda < V_{eff} (\alpha \phi \rightarrow \infty )= C
\end{equation}
where $C = \frac{f_1^2}{4\,f_2(1+\kappa^2\epsilon f_1^2/f_2)}+ \Lambda =
\frac{f_1^2}{4f_2}\,A + \Lambda $.
notice that we assume above that $f_1>0$ and $M < 0$, but $f_1<0$ and $M > 0$ would be indistinguishable from that situation, that is, the important requirement is $f_1/M < 0$.
We could  have a scenario where we start the non singular emergent universe at $ \phi \rightarrow \infty $ where
$V_{eff} (\alpha \phi \rightarrow \infty )= \frac{f_{1}^{2}}{4(\epsilon \kappa ^{2} f_{1}^{2} + f_{2})}+\Lambda$, which then slow rolls, then inflates \cite{SIchile} and finally gets trapped in the local minimum with energy density $V_{eff}(min, M < 0)= \Lambda$, that was the picture favored in
\cite{SIchile}, while here we want to argue that the most attractive and relevant description for the final state of our Universe is realized after inflation in the flat region 
$\phi \rightarrow -\infty$, since in this region the vacuum energy density is positive and bounded from above, so its a good candidate for our present state of the Universe. It remains to be  seen howevere whether a smooth transition all the way from  $\phi \rightarrow \infty$ to 
$\phi \rightarrow -\infty$ is possible.

In order to discuss the possiblility of transition to $\phi \rightarrow -\infty$ . In our case, since we are interested in a local minimum 
between $\phi \rightarrow \infty$ or $\phi \rightarrow -\infty$, we can take $M$ of either sign. 

Taking for definitness $f_1>0$, $f_2>0$, $A>0$, $\epsilon<0$, we see that there will be a point, defined by 
$\epsilon \kappa ^{2}(f_{1}e^{\alpha \phi}  +  M )^{2} + f_{2}e^{2 \alpha \phi } =0$  
where the effective potential for a constant field $\phi$, then   $V_{eff}$ as given by (\ref{effppluslambda}),  will spike to $\infty$, go then down to $-\infty$ and then asymptotically approach its possitive asymptotic value at $\phi \rightarrow -\infty$. This has the appearence of a potential barrier. However, this is deceptive, such barrier may exist for constant $\phi$, but can be avoided by considering 
time dependence, say for no space dependence and $\dot{\phi}^2$ given by

\begin{equation}
\dot{\phi}^2 = - \frac{1}{\epsilon \kappa ^{2}}
\end{equation}

which has a solution in the real domain for $\epsilon < 0$. For this case $R$ (which is not a Riemannian curvature), as given by (\ref{e51}) diverges. In this case then

\begin{equation}
V_{eff}  = \frac{\epsilon R^{2} + U}{(\chi -2 \kappa \epsilon R)^{2} }+ \Lambda \rightarrow  \frac{1}{4\epsilon\kappa^{2}} + \Lambda = \Delta \lambda
\end{equation}
  that is, for this value of $\dot{\phi}^2$, regardless of the value of the scalar field, the value of $V_{eff}$ becomes degenerate with its value for constant and arbitrarily negative $\phi$, which is our candidate vacuum for the present state of the Universe. Therefore there is no barrier that prevents from us reaching arbitrarily negative $\phi$ from any point in field space in this model.

\section{Discussion, the creation of the universe as a "threshold event" for zero present vacuum energy density}

We have considered a non singular origin for the Universe starting from an Einstein static Universe, the so called "emergent universe" scenario, in the framework of a theory which uses two volume elements  $\sqrt{-{g}}d^{4}x$ and $\Phi d^{4}x$, where $\Phi $ is a metric independent density, used as an additional measure of integration. Also curvature, curvature square terms and for scale invariance a dilaton field $\phi$  are considered in the action. The first order formalism was applied.  The integration of the equations of motion associated with the
new measure gives rise to the spontaneous symmetry breaking (S.S.B) of scale invariance (S.I.).
After S.S.B. of S.I., using the the Einstein frame metric, it is found that a non trivial potential for the dilaton is generated. 
One could question the use of the Einstein frame metric $\overline{g}_{\mu\nu} $ in contrast to the original metric $g_{\mu\nu} $ . In this respect, 
it is interesting to see the role of both the original metric and that of the Einstein frame metric
in a canonical approach to the first order formalism. Here we see that the original metric does not have a canonically conjugated momentum (this turns out to be zero), in contrast, the canonically conjugated momentum to the conection turns out to be a function exclusively of 
$\overline{g}_{\mu\nu} $, this Einstein metric is therefore a genuine dynamical canonical variable, as opposed to the original metric. 

There is also a lagrangian formulation of the theory which uses $\overline{g}_{\mu\nu} $, what we can call the action in the Einstein frame. In this frame we can quantize the theory for example and consider contributions without reference to the original frame, thus possibly considering breaking the TMT structure of the theory, but such breaking will be done "softly" through the introduction of a cosmological term only. In previous studies, we have found that the TMT structure of the theory, where neither the lagrangian $L_1$that couples to $\sqrt {-g}$,  or $L_2$, that couples to $\Phi$ depend on the measure fields, is protected by an infinite dimensional symmetry $\varphi_a \rightarrow \varphi_a + f_a(L_2)$, where $f_a(L_2)$ is an arbitrary function of $L_2$. The additional cosmological term, introduced here in the Einstein frame, does not have a representation of this form in the original frame, therefore breaking the TMT structure (therefore the infinite dimensional symmetry would be also broken by quantum effects). Surpringly, the remaining terms of the theory, reminiscent from the original TMT structure will be enough
to control the strength of this additional cosmological term once we demand that the universe originated from a non singular and stable emergent state.

In the Einstein frame we argue that the cosmological term  parametrizes the zero point fluctuations.

The resulting effective potential for the dilaton contains two flat regions, for 
$\phi \rightarrow \infty$ relevant for the non singular origin of the Universe, followed by an inflationary phase and then transition to $\phi \rightarrow -\infty$, which in this paper we take as describing our present Universe. An intermediate local minimum is obtained if $f_1/M<0$, the region as $\phi \rightarrow \infty$ has a higher energy density than this local minimum and of course of the region $\phi \rightarrow -\infty$, if  $A>0$ and $f_2 >0$. $A>0$ is also required for satisfactory slow roll in the inflationary region $\phi \rightarrow \infty$ (after the emergent phase). The dynamics of the scalar field becomes non linear and these non linearities
are instrumental in the stability of some of the emergent universe solutions, which exists for a parameter range
of values of the vacuum energy in $\phi \rightarrow -\infty$, which must be positive but not very big, avoiding the extreme fine tuning required to keep the vacuum energy density of the present universe small. A sort of solution of the Cosmological Constant Problem, where an a priori arbitrary cosmological term is restricted by the consideration of the nonsingular and stable emergent origin for the universe.

Notice then that the creation of the universe can be considered as a "threshold event" for zero present vacuum energy  density, that is a threshold event for  $\Delta \lambda = 0$ and we can learn what we can expect in this case by comparing with well known threshold events. For example in particle physics, the process $e^{+}+e^{-} \rightarrow \mu^{+}+\mu^{-}$, has a cross section of the form (ignoring the mass of the electron and considering the center of mass frame, $E$ being the center of mass energy of each of the colliding $e^{+}$ or $e^{-}$), 

\begin{equation}
 \sigma_{e^{+}+e^{-} \rightarrow \mu^{+}+\mu^{-}} = \frac{\pi \alpha^{2}}{6 E^{2} }\left[ 2 + \frac{m_{\mu}^2}{E^2}  \right]\sqrt{\frac{E^{2}-m_{\mu}^2}{ E^{2}}}
\end{equation}
for $E>m_{\mu}$ and exactly zero for $E < m_{\mu}$ . We see that exactly at threshold this cross section is zero, but at this exact point it has a cusp, the derivative is infinite and the function jumps as we slightly increase $E$.
By analogy, assuming that the vacuum energy can be tuned somehow (like the center of mass energy $E$ of each of the colliding particles in the case of the annihilation process above), we can expect zero probability for exactly zero vacuum energy density $\Delta\lambda =0$, but that  soon after we build up any positive $\Delta\lambda$ we will then able to create the universe, naturally then, there will be a creation process resulting in a universe with a small but positive $\Delta\lambda$ which represents the total energy density for the region describing the present universe, $\phi \rightarrow -\infty$.

One challenge would be to in fact calculate from this approach the probability of creating the universe with a given vacuum energy density of the vacuum for the region describing the present universe, $\phi \rightarrow -\infty$, the same way we calculate the probability of the process $e^{+}+e^{-} \rightarrow \mu^{+}+\mu^{-}$. This will give us the probability of a given present vacuum energy density.

\section{Acknowledgements}

I would like to thank  Sergio del Campo, Ramon Herrera, Alexander Kaganovich and Pedro Labrana for very important discussions and to the Physics Department of the Pontificia Universidad de Valparaiso, Chile for hospitality during several visits.

\break

\end{document}